\documentclass[twocolumn,10pt]{asme2ej}

\usepackage{amsmath}
\usepackage{caption}
\usepackage{amsfonts}
\usepackage[export]{adjustbox}
\usepackage{listings}
\usepackage{hyperref}

\hypersetup{
	colorlinks=true,
	linkcolor=blue,
	citecolor=blue,
	urlcolor=blue,
}

\newcommand{\vc}[1]{\boldsymbol{#1}}

\title{svMorph: Interactive geometry-editing tools for virtual patient-specific vascular anatomies}

\author{Jonathan Pham
    \affiliation{   Graduate Research Assistant\\
                    Department of Mechanical Engineering\\
                	Stanford University\\
                	Stanford, California 94305
    }
}

\author{Sofia Wyetzner
    \affiliation{   Graduate Research Assistant\\
                	Department of Computer Science\\
                	Stanford University\\
                	Stanford, California 94305
    }
}

\author{Martin R. Pfaller
    \affiliation{   Postdoctoral Scholar\\
                	Department of Pediatrics\\
                	Stanford University\\
                	Stanford, California 94305
    }
}

\author{David W. Parker
    \affiliation{   Research Software Engineer\\
                	Stanford Research Computing Center\\
                	Stanford University\\
                	Stanford, California 94305
    }
}

\author{Doug L. James
    \affiliation{   Professor\\
                	Department of Computer Science\\
                	Stanford University\\
                	Stanford, California 94305
    }
}

\author{Alison L. Marsden
    \affiliation{   Professor\\
                	Department of Bioengineering\\
                	Department of Pediatrics\\
                	Stanford University\\
                	Stanford, California 94305
    }
}

\begin{document}

\maketitle    

\begin{abstract}
    {\it We propose svMorph, a framework for interactive virtual sculpting of patient-specific vascular anatomic models. Our framework includes three tools for the creation of tortuosity, aneurysms, and stenoses in tubular vascular geometries. These shape edits are performed via geometric operations on the surface mesh and vessel centerline curves of the input model. The tortuosity tool also uses the physics-based Oriented Particles method, coupled with linear blend skinning, to achieve smooth, elastic-like deformations. Our tools can be applied separately or in combination to produce simulation-suitable morphed models. They are also compatible with popular vascular modeling software, such as SimVascular. To illustrate our tools, we morph several image-based, patient-specific models to create a range of shape changes and simulate the resulting hemodynamics via three-dimensional, computational fluid dynamics. We also demonstrate the ability to quickly estimate the hemodynamic effects of the shape changes via automated generation of associated zero-dimensional lumped-parameter models.}
\end{abstract}

\section{Introduction}

    Computational fluid dynamics (CFD) simulations in patient-specific vascular anatomies are increasingly used to support applications in surgical and treatment planning and medical device design for cardiovascular disease in children and adults \cite{Ballarin2016, trusty2019, Kasinpila2021, Lan2022, Pant2022}. Simulations can also elucidate the relationship between biomechanics and vascular disease progression by characterizing mechanical stimuli that can be difficult to obtain experimentally \cite{Vedula2017, Schwarz2021, Dong2021, Teeraratkul2021}. However, the patient-specific modeling pipeline can be a tedious process, hindered by three primary bottlenecks. First, building an initial model and mesh of a patient-specific vascular anatomy is a laborious task that typically requires careful creation of vessel centerline paths and segmentations of the corresponding lumen from imaging data \cite{Taylor2009, Updegrove2017}. These tasks routinely require expert user knowledge in medical image interpretation to correctly position the pathlines and the segmentations, as well as many hours of manual positioning and correction. Second, modifying the geometry of existing models to study the hemodynamic effects of different treatment options or degrees of disease severity typically requires the user to perform tedious by-hand manipulation of pathlines and segmentations before regenerating the entire model. While this may be acceptable for a limited number of simpler scenarios, it quickly becomes intractable for complex cases with numerous possible anatomic changes.  With by-hand methods, it is also often difficult to ensure a smooth and watertight model needed for high-quality simulations. Third, performing three-dimensional (3D) CFD simulations in a given model often requires hours to days of runtime on a multi-processor supercomputer.
    
    There have been recent significant strides towards addressing these challenges. First, to address bottlenecks in image segmentation, Maher et al. developed regression-based neural networks to quickly and accurately segment vascular lumen contours, significantly reducing model generation time \cite{Maher2019, Maher2020}. Other groups have also demonstrated success with automated segmentation methods in cardiac anatomies and coronary artery disease \cite{Kong2020, Schaap2011}.
    
    Second, several frameworks have been proposed to tackle the model-modification challenge. The SURGEM tool was developed to support surgical planning for congenital heart diseases, such as single ventricle physiology, via interactive shape editing of pediatric cardiovascular anatomies \cite{Pekkan2008, Luffel2016}. To morph models in SURGEM, users utilize handheld, six-degree-of-freedom motion trackers or styluses with tablets. Such hardware is not typically readily available to clinicians or cardiovascular simulation practitioners, however. There also remains a need for seamless transition to generating CFD simulation results from the edited geometries in this framework.
    Recently, the Harvis framework was also developed to enable hands-on anatomic modification of aortic models in virtual reality for medical training applications \cite{Shi2020}. To enable script-based editing, the morphMan framework was created \cite{Kjeldsberg2019}. This tool enables command-line-based shape manipulation of vascular models. However, geometric changes in morphMan are performed by converting the input surface model into an equivalent Voronoi diagram for editing. This requires a reverse conversion to obtain the morphed mesh, which may have a different nodal connectivity from the input mesh. This may preclude direct use of the boundary conditions previously prescribed to the input mesh nodes.
    
    Third, to address bottlenecks in simulation time, data-driven techniques via model order reduction and deep learning have been applied to accelerate 3D CFD simulations in vascular anatomies \cite{Pegolotti2021, Liang2020}. Pfaller et al. also developed techniques to initialize 3D CFD simulations from reduced-order models, reducing the number of timesteps needed to obtain converged results \cite{Pfaller2021}. These reduced-order models were generated from their respective 3D anatomic models via automated methods \cite{Pfaller2022}.
    
    Focusing on the challenges of model morphing, we build on the previous efforts to develop svMorph, a virtual framework for interactive sculpting of patient-specific anatomies. In contrast to morphMan, the morphing tools in svMorph operate directly on the input surface mesh and model centerline. This circumvents the need for auxiliary model representations, reducing computational complexity. Furthermore, the use of the surface mesh and centerlines for morphing makes our sculpting tools directly compatible with open-source vascular modeling software, such as SimVascular \cite{Updegrove2017}, providing a seamless transition from modeling to simulation. Our framework was also developed to be suitable for a wide range of vascular anatomies, rather than being targeted primarily at congenital heart diseases and aortic physiologies, as in the case of SURGEM and Harvis. The morphing tools can also be applied to a wide range of vascular diseases and their use requires only a keyboard and a mouse. We demonstrate these capabilities on three patient-specific models and show that our sculpting tools yield analysis-suitable, watertight meshes by performing patient-specific 3D CFD simulations on several morphed geometries. Unlike previous frameworks, svMorph is also integrated with a reduced-order 0D solver to demonstrate the capability for fast estimation of hemodynamic changes resulting from geometric edits. Our framework is intended to support surgical and treatment planning needs, in which it is desirable to rapidly explore shape manipulations and resulting hemodynamics. We note that the model editing tools in svMorph may also hold value in future shape optimization and uncertainty quantification efforts, in which numerous design instances must be explored in an automated manner \cite{Dur2011, Maher2021}. We also note that other reduced-order modeling methods could be interfaced with svMorph in future work.
    
    \begin{figure}
        \centering
        \includegraphics[width = 0.41\textwidth, height = \textheight, keepaspectratio]{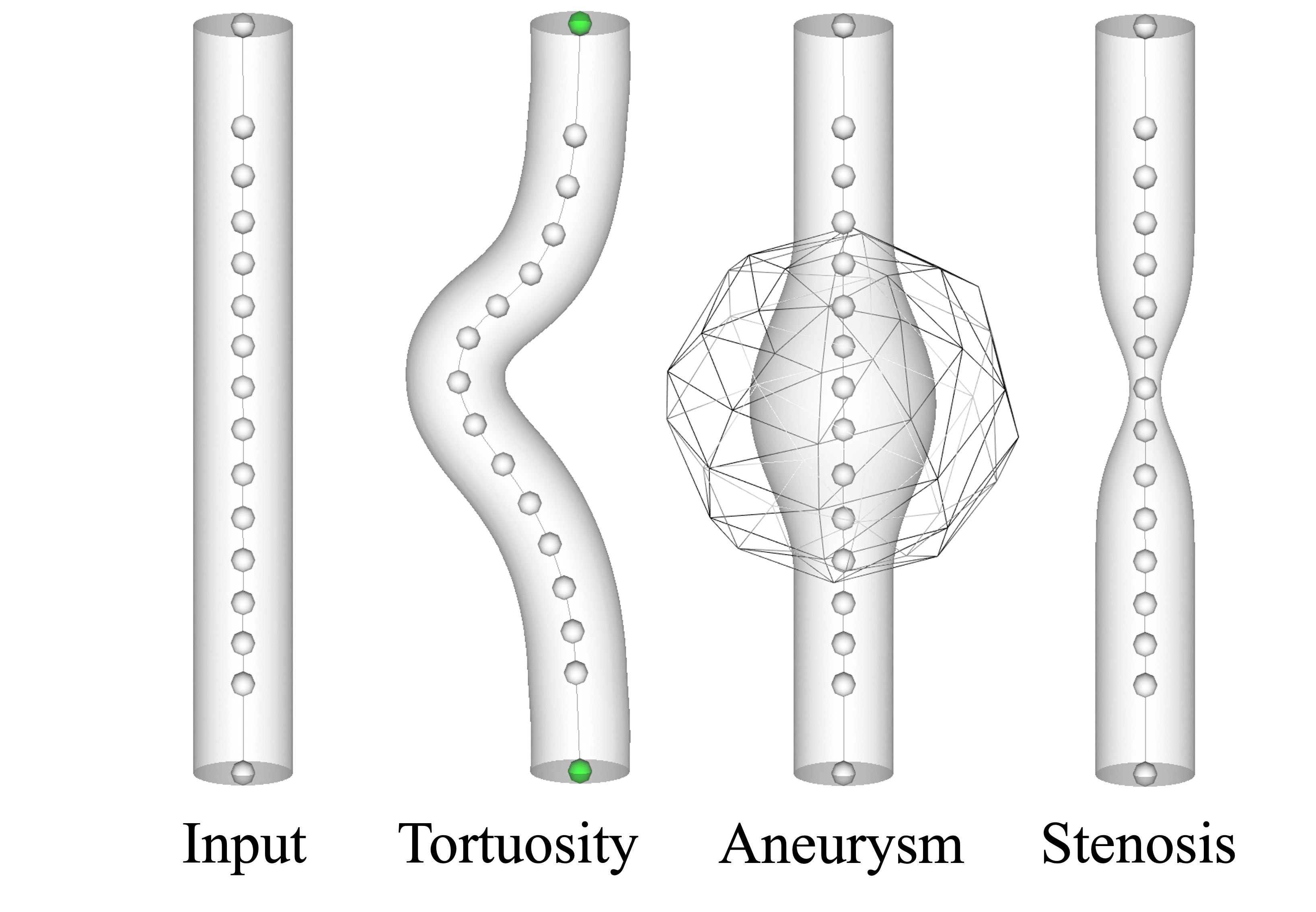}
        \caption{We illustrate our tortuosity, aneurysm, and stenosis tools on an idealized vessel. The green control points shown in the tortuosity case are pinned in our Oriented Particles simulation. Our aneurysm tool uses a radius-of-influence to define the aneurysm location. \label{fig_svMorph_tools}}
    \end{figure}
    
    \begin{figure}
        \centering
        \includegraphics[width = 0.25\textwidth, height = \textheight, keepaspectratio]{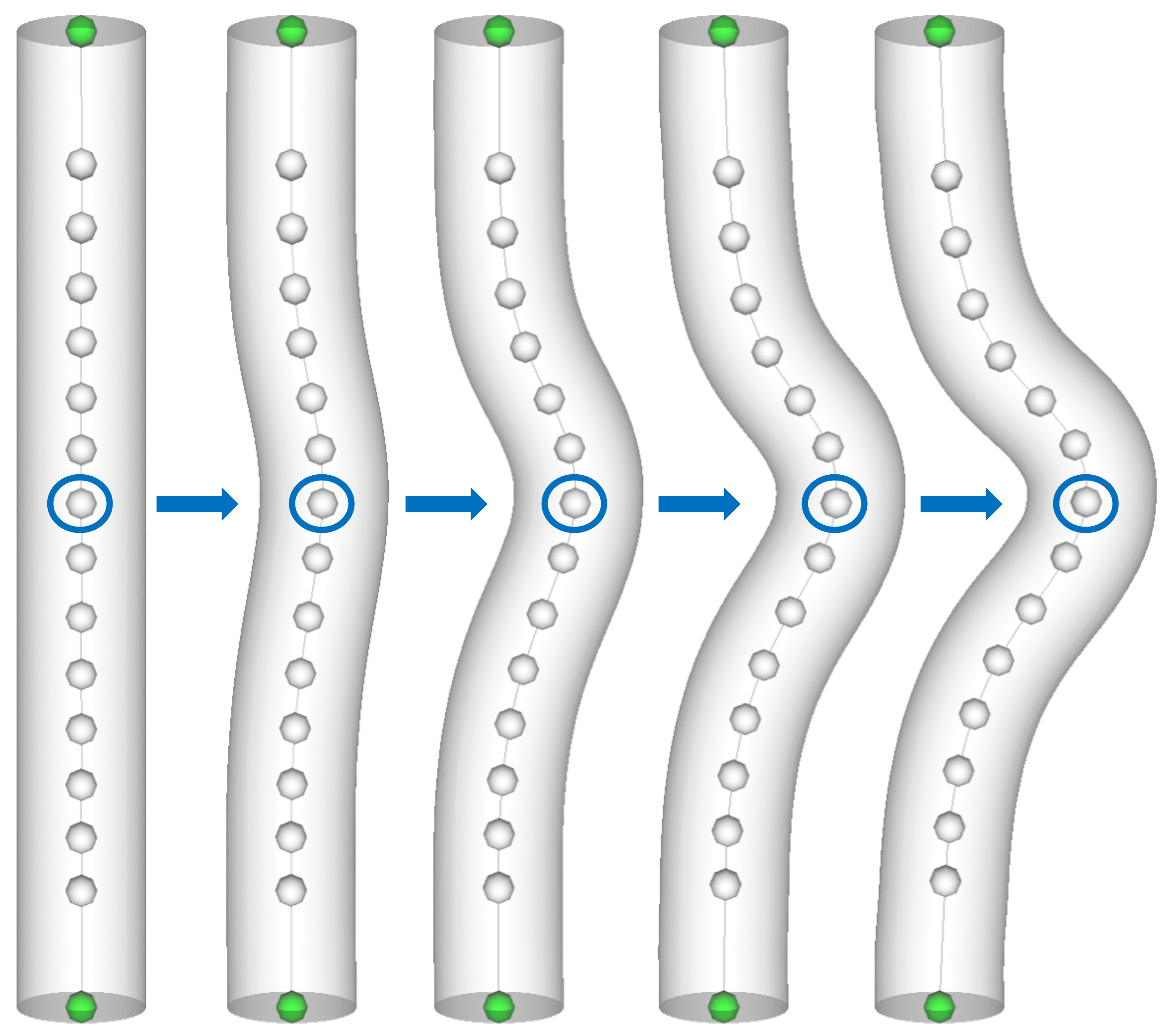}
        \caption{The centerline control point circled in blue is selected with the mouse and dragged over to create tortuosity in an idealized blood vessel. The dynamic deformation is shown at progressive time steps in the OP simulation. \label{fig_tortuosity_progression}}
    \end{figure}
    
\section{Methods}
    
    We use the Visualization Toolkit library (VTK) for svMorph's graphics engine \cite{Schroeder2000}. Three interactive shape-editing tools for tubular vascular geometries are currently implemented in our framework: tortuosity inducement, aneurysm generation, and stenosis creation. Figure \ref{fig_svMorph_tools} shows examples of these tools. Each tool can be applied independently to create features of interest. The tools can also be used in combination by applying each tool consecutively, one after the other, to create more complex shapes, as will be shown in a later example.

    Geometry editing operations in svMorph directly modify the centerlines and the triangulated surface mesh of the input model, both of which are represented with the VTK's \texttt{vtkPolyData} class \cite{Schroeder2000} (Fig.~\ref{fig_svMorph_tools}). To morph the geometry, users manipulate the positions of the centerline control points. These changes are then propagated to the surface mesh to yield the deformed vessel surface. We provide details on the morphing methods in the sections below.
    
    Users can export the edited centerlines and surface mesh to perform CFD simulations or reduced-order modeling in the morphed anatomies. In this work, we use SimVascular, an open-source cardiovascular modeling software, to generate simulation files from the svMorph geometry data \cite{Updegrove2017}. If a 3D CFD simulation is desired, volume meshing can readily be performed via SimVascular using its native meshing tool. No preprocessing is required to convert the surface mesh into an equivalent volume mesh, which is represented with VTK's \texttt{vtkUnstructuredGrid} class, in SimVascular. For 0D simulations, the modified centerline geometry can be used directly to generate the simulation solver files.
    
    \subsection{Tortuosity tool}
        
        Our tortuosity tool is a simulation-driven deformer that enables smooth curvature inducement in vascular models. This tool consists of two components: a simulation step that deforms the centerlines and a skinning step that propagates the centerline deformations to the surface mesh. We describe the simulation method below. The skinning step is discussed immediately afterwards.
        
        \begin{figure}
            \centering
            \includegraphics[width = 3.25in, height = \textheight, keepaspectratio]{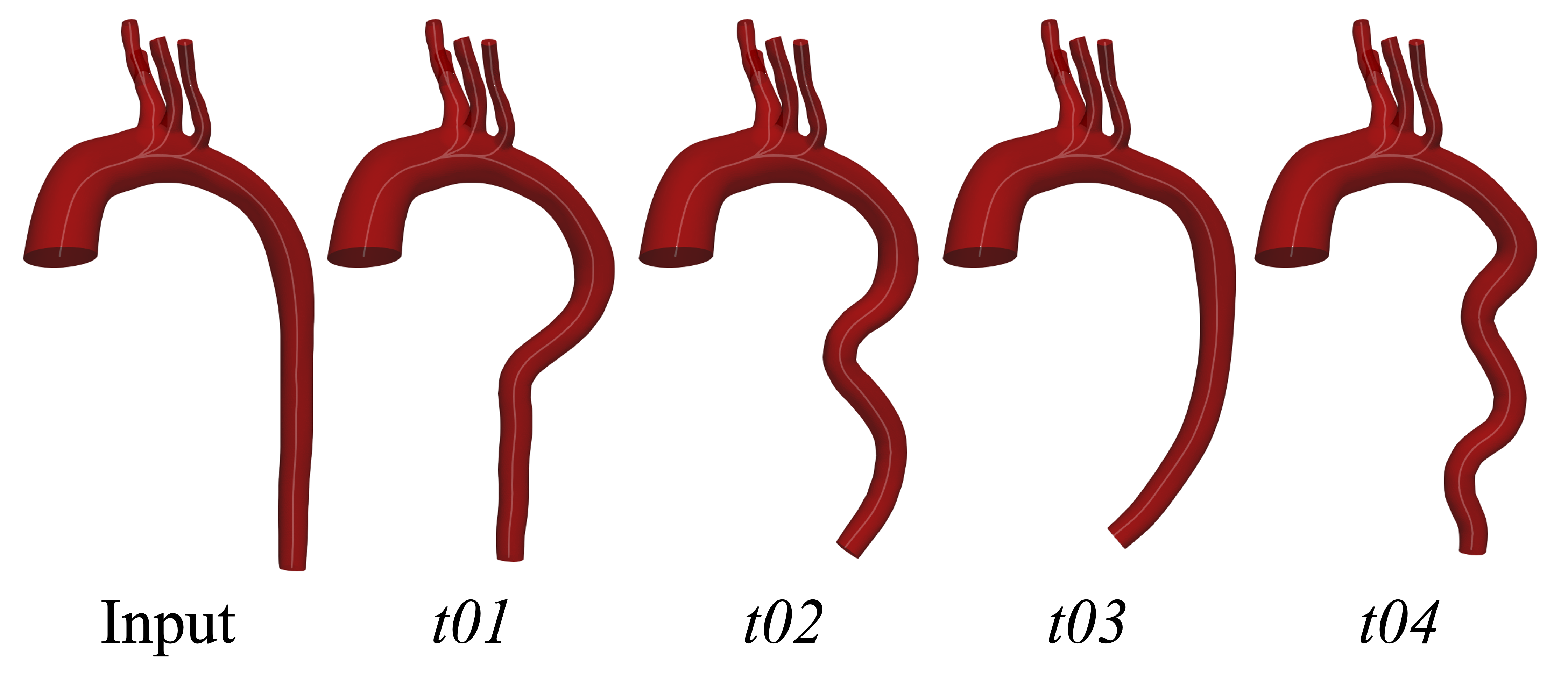}
            \caption{We apply our tortuosity tool to create several curvature shapes on a patient-specific aorta\label{fig_tortuosity_examples}}
        \end{figure}
    
        To use the tortuosity tool, the user pins a set of control points on the centerlines. These points will not deform during the simulation. Instead, they act as spatial anchors, as shown by the green points in Fig.~\ref{fig_svMorph_tools}. By default, all control points located at the junctions between branches on the centerlines are pinned. The user then selects an unpinned control point with their mouse in the graphical user interface (GUI) of svMorph and drags it to a new location. Throughout the entire dragging event, the positions and orientations of all other unpinned control points are simultaneously updated to yield a smooth, tortuous deformation of the centerlines via the Oriented Particles (OP) method, a physics-based technique developed by the computer graphics community for simulating the dynamic deformation of solids \cite{Muller2011}. 
        Note that each incremental mouse motion in the dragging event corresponds to a single time step in the OP simulation step. This allows users to see the centerline dynamically deform in the GUI, as exemplified in an idealized blood vessel in Fig.~\ref{fig_tortuosity_progression}.
        
        The OP method updates the state of each control point by iteratively correcting an initial prediction of its deformed state subject to shape-matching constraints \cite{Muller2004}. The initial prediction is calculated from the current positions and velocities of the control points. We note that in the first iteration of this simulation, the positions of the control points are initialized to their default positions on the centerlines and the velocities are initialized to zero. To satisfy the shape-matching constraints, we compute a goal position for each unpinned control point. This is computed by finding the transformation matrix for the current control point and its two adjacent control points that effectively minimizes the difference between the morphed state and rest state of the region in a least-squares sense. We refer interested readers to the shape-matching paper of M$\ddot{u}$ller et al. for details on the computation of this transformation matrix \cite{Muller2004}. After computing the goal position, we correct the deformed position of each control point to be a weighted sum of its goal position and the initial prediction. This weighted sum allows us to model elastic-like deformations in the centerlines. Following the iterative shape-matching correction step, the OP method then uses the final and initial position and orientation of each control point, as well as the time step size, to update its velocity. This update is done to ensure consistency with the final deformed state of the control points.

        \begin{figure*}
            \centering
            \includegraphics[width = \textwidth, height = \textheight, keepaspectratio]{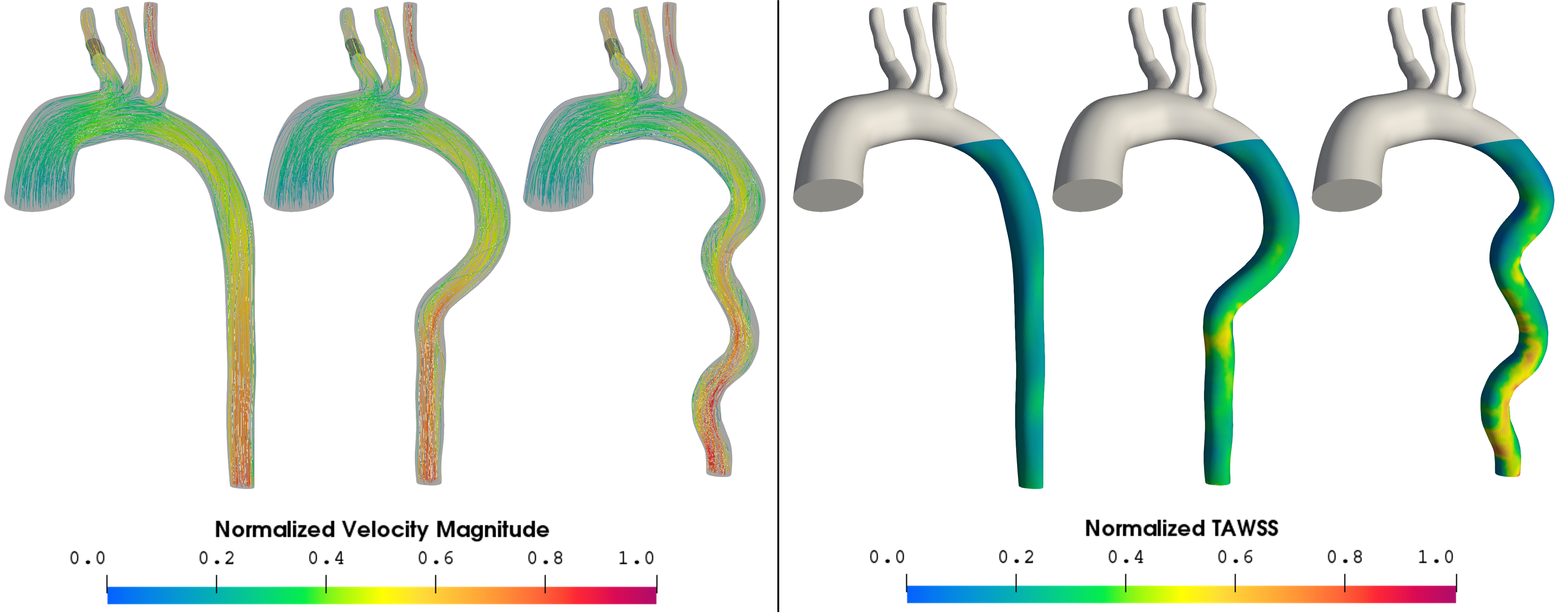}
            \caption{The 3D-simulated velocity streamlines (left panel) and the TAWSS (right panel), normalized by their respective maximum values, are compared between the input aorta and two tortuous shapes. We see that increasing the descending aorta curvature leads to increased TAWSS. \label{fig_aorta_results}}
        \end{figure*}
        
        Following OP-based deformation of the centerline, we use linear blend skinning to deform the surface mesh \cite{Kavan2014}. With this method, a deformable object is composed of a skeleton and a surface mesh. The skeleton consists of the centerline control points in svMorph. Each surface mesh vertex is attached to the set of bones, i.e., control points, closest to it on the skeleton. When a bone rotates or translates, the attached surface mesh vertices transform correspondingly in linear blend skinning. The position of each surface mesh vertex is transformed according to a weighted sum of the transformation matrices corresponding to its $n_{b}$ bones. The deformed position, $\vc{x}_{i}'$, of a surface mesh vertex, $i$, whose rest position is $\vc{x}_{i}$, is computed via $\vc{x}_{i}' = \sum_{j=1}^{n_{b}} w_{i, j}\vc{T}_{j}\vc{x}_{i}$,
        where $\vc{T}_{j}$ is the transformation matrix of bone $j$, and $w_{i, j}$ is the weight associated between vertex $i$ and bone $j$. For each vertex, the weights across its bones satisfy a partition of unity to ensure translationally invariant deformations \cite{Jacobson2015}. We use an exponentiated inverse distance for our weights, $w_{i, j} = \left(\frac{1}{d}\right)^{s}$, where $d$ is the distance between vertex $i$ and centerline control point $j$; $s$ is a user-defined exponent. To ensure the cap faces on the surface mesh stay planar during deformation, the cap faces vertices each use only one bone.
        
    \subsection{Aneurysm tool}
        
        Our aneurysm tool allows users to interactively create aneurysms on patient-specific vascular models. An aneurysm is generated on the surface mesh of the input model according to the following procedure. First, the user selects a control point on the centerlines to define the center position, $\vc{c}$, of the aneurysm. A spherical zone-of-influence is centered here. This region defines the neighborhood in which the aneurysm will be created, as depicted in Fig.~\ref{fig_svMorph_tools}. The user can vary the radius, $R$, of this sphere to change the size of the aneurysm generated. Each vertex, $\vc{x}$, on the surface mesh is displaced in its normal direction by a displacement magnitude of $d\left(\vc{x}\right) = \left(1 - \frac{||\vc{x} - \vc{c}||_{2}}{R}\right)^{\lambda}$
        to automatically generate the aneurysm, where $\lambda$ is a user-defined exponent that controls the falloff rate of the aneurysm. With this displacement function, the vertices on the surface further away from the center of the aneurysm are displaced less. After the aneurysm is created on the surface, the cross-sectional areas of the lumen on the centerline model are updated. We note that in this tool, the aneurysm is created by expanding the vessel lumen without accounting for possible contributions from intraluminal thrombus that may be clinically relevant in some cases.
        
    \subsection{Stenosis tool}
    
        Our stenosis tool allows users to contract the cross-sectional area of the lumen in a vessel to create stenoses of a specified size. To use this tool, the user first selects a control point on the centerline. The stenosis will be centered at the position, $\vc{c}$, of this point. The user then specifies the desired severity of the stenosis as a percentage of the original cross-sectional lumen area. The stenosis is then automatically generated on the surface mesh through a displacement of its vertices according to an analytical shape profile. In this work, we represent stenoses with the Gaussian profile, $d\left(\vc{x}_{proj}\right) = e^{-\left(\frac{||\vc{x}_{proj} - \vc{c}||_{2}}{\sigma}\right)^{2}}$, where
        $\vc{x}_{proj}$ is the projection of a surface mesh vertex onto the centerline. However, we note that our method is agnostic to the shape profile used. Other analytical functions, such as cosine curves, can be used as well.
        
        The vector pointing from the vertex to its projection defines the direction in which to displace the vertex. The Gaussian profile computes the normalized displacement magnitude for each vertex. It is evaluated as a function of distance of the projected point from the center of the stenosis along the centerline axis. This field has an amplitude of one at the center of the stenosis, but it exponentially decays to zero according to its width, $\sigma$, another user-controlled parameter. The normalized displacement magnitude computed by this function is then multiplied by the previously specified degree of severity, along with the direction of displacement, to generate the stenosis on the surface mesh. Through this approach, the cross-sectional areas of the lumen stored on the centerline model are also automatically updated.
    
    \subsection{Three-dimensional hemodynamic simulations}
        The hemodynamics of blood vessels is governed by the 3D incompressible Navier-Stokes equations, 
        \begin{align}
            \frac{\partial \vc{u}}{\partial t} + \vc{u} \cdot \nabla \vc{u} &= -\frac{1}{\rho}\nabla p + \nu \nabla^{2} \vc{u} + \vc{b},  \label{eqn_3d_momentum}\\
            \nabla \cdot \vc{u} &= 0. \label{eqn_3d_mass}
        \end{align}
        Here, $\vc{u}$ is the velocity of the blood, $p$ is the pressure, and $\vc{b}$ is the body force; $\rho$ and $\nu$ are the density and kinematic viscosity of the blood, respectively. Equations \eqref{eqn_3d_momentum} and \eqref{eqn_3d_mass} represent the conservation of momentum and mass, respectively, in the system. We couple this system of equations to a set of inlet and outlet boundary conditions, represented as lumped-parameter networks and differential-algebraic equations, to model patient-specific hemodynamics. Here, we use three-element Windkessel models, inflows, no-slip conditions on the walls, and rigid-wall assumptions as our boundary conditions \cite{vignonclementel06}.
        
        We numerically discretize the Navier-Stokes equations using a stabilized finite element method with P1-P1 elements for spatial discretization and the second-order generalized-$\alpha$ method for time advancement \cite{franca92, whiting01}. We solve the resulting discretized system, coupled to the lumped-parameter boundary conditions, with a customized linear solver and preconditioner and a modular coupling method \cite{moghadam13, esmailymoghadam13}. SimVascular's 3D solver, \texttt{svSolver}, is available open-source at \url{https://github.com/SimVascular/svSolver}.
        \begin{figure}
            \centering
            \includegraphics[width = 3.25in, height = \textheight, keepaspectratio]{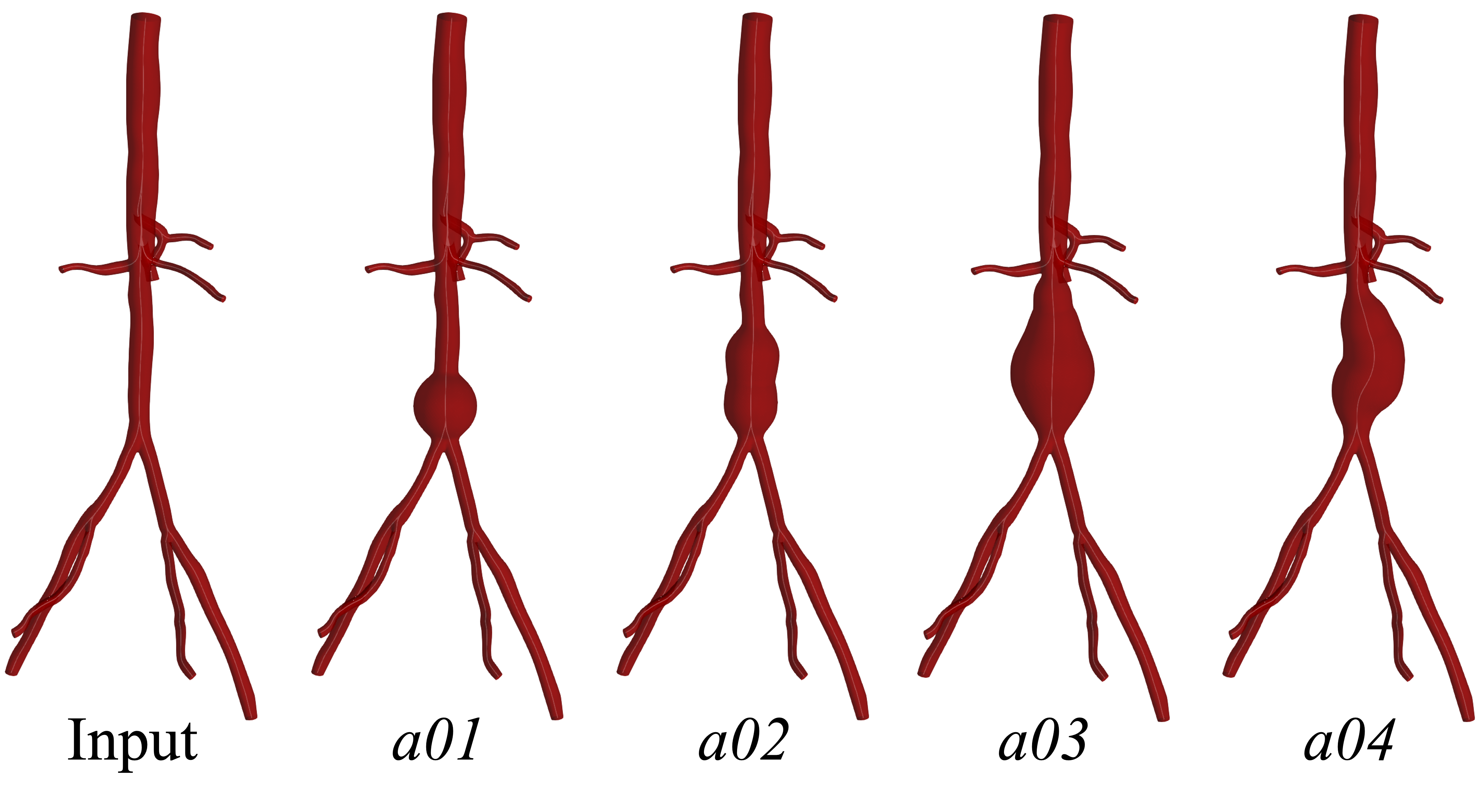}
            \caption{We create aneurysms of various shapes for an aortofemoral model using our aneurysm tool. In \textit{a04}, we combine the shape edits generated by our tortuosity and aneurysm tools to create the asymmetric aneurysm. \label{fig_aneurysm_examples}}
        \end{figure}

    \subsection{Zero-dimensional hemodynamics simulations}
        \begin{figure*}
            \centering
            \includegraphics[width = \textwidth, height = \textheight, keepaspectratio]{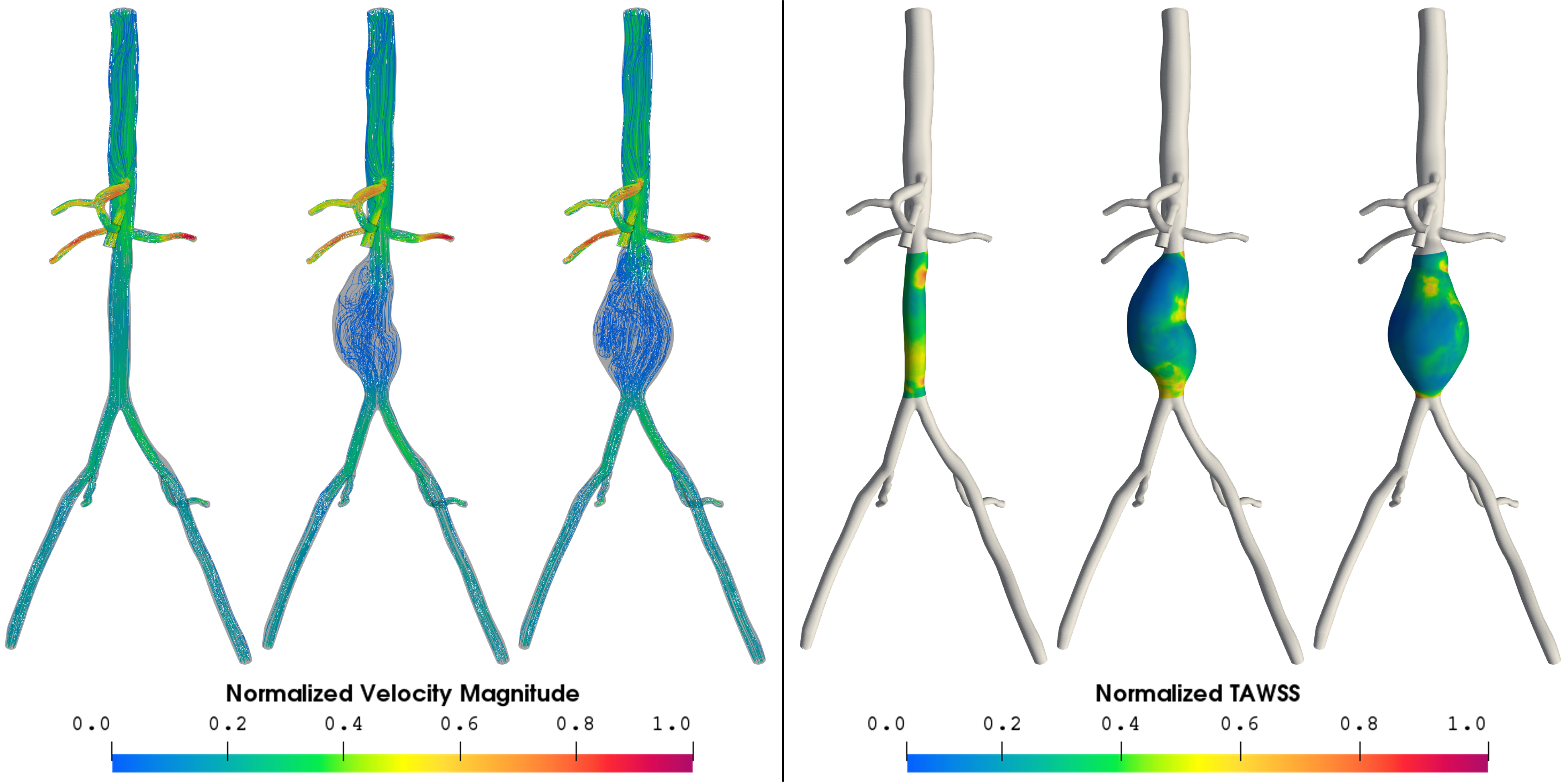}
            \caption{We compare the 3D velocity streamlines (left panel) and TAWSS (right panel) between the input aortofemoral model and two aneurysmal shapes. The velocity magnitude and the TAWSS were normalized by their respective maximum values. The aneurysms have internal flow recirculation and decreased TAWSS. \label{fig_aortofemoral_results}}
        \end{figure*}
        Three-dimensional solutions of the Navier-Stokes equations offer insights into local hemodynamic patterns. However, solving these equations is computationally expensive, often requiring hours of runtime on a high-performance computing cluster. A popular, alternative strategy for vascular simulations is the low-fidelity 0D lumped-parameter model, which requires only seconds to minutes on a single processor. Previous works have also shown that 0D models can often accurately simulate bulk hemodynamics when compared against their 3D counterparts \cite{Pfaller2022}. Despite their simplicity, 0D models have produced valuable insights in the context of surgical planning for congenital heart disease and other applications \cite{Conover2018}. The drawback of 0D models, however, is that they are only capable of simulating bulk hemodynamic quantities, such as flow rates and cross-sectionally averaged pressures. Despite this drawback, 0D models are often adequate surrogates for 3D simulations if global hemodynamics are the primary interest, as is often the case in clinical settings. We note that the accuracy of 0D models is known to break down in the setting of diseased anatomies, flow separation, and flows with strong nonlinearity; these limitations are well documented in other studies \cite{Pfaller2022}. Performance can be improved with the use of empirical loss models in relevant cases.
        
        Lumped-parameter 0D models are composed of circuit-like elements: resistors, inductors, and capacitors. In the vascular modeling context, these elements model model viscous losses, flow inertia, and vascular wall compliance, respectively. Each element is governed by its own local equations. When we connect the individual elements together with the boundary conditions, we obtain 
        \begin{equation}
            \vc{E}\left(\vc{y}, t\right) \cdot \frac{d\vc{y}}{dt} + \vc{F}\left(\vc{y}, t\right)\cdot \vc{y} + \vc{c}\left(\vc{y}, t\right) = \vc{0}, \label{eqn_0d_dae}
        \end{equation}
        a system of differential-algebraic equations. This system governs the hemodynamics in the entire 0D model. Here, $\vc{y}$ is a vector composed of the flow rates and pressures, while $\vc{E}$, $\vc{F}$, and $\vc{c}$ contain the coefficients parameterizing the lumped-parameter elements used in the 0D model. We use SimVascular to generate the 0D model from the morphed centerline using the automated pipeline developed by Pfaller et. al \cite{Pfaller2022}. We use the open-source \texttt{svZeroDSolver} code (\url{https://github.com/SimVascular/svZeroDSolver}) to numerically solve Eqn.~\eqref{eqn_0d_dae} with the generalized-$\alpha$ method \cite{jansen00}.
        
\section{Results}
    
    We illustrate the morphing tools in svMorph on three patient-specific vascular models: an aorta, an aortofemoral network, and an aorta-coronary system. We obtained each model from the Vascular Model Repository (\url{https://vascularmodel.com}) \cite{Wilson2013}.
    
    We demonstrate that our morphing tools yield analysis-suitable meshes by simulating the hemodynamics in the morphed aorta and aortofemoral models. To generate the volume mesh needed for 3D CFD analysis of these models, we used SimVascular's native volume meshing tools. We also leverage 0D models to obtain improved initial conditions for our 3D simulations to reduce the number of time steps that must be simulated to reach a limit state \cite{Pfaller2021}. We additionally demonstrate in the coronary physiology that we can quickly estimate the global hemodynamic changes resulting from the deformation via 0D simulations. We simulate all models until the inlet and outlet flow rates and pressures have converged to a time-periodic limit state, which we define as a 1\% maximum difference between adjacent cardiac cycles.
    
    \subsection{Inducing curvature in an aorta}
        
        Previous studies have indicated that aortic tortuosity tends to increase with age \cite{Belvroy2019, Ciurica2019}. However, it is still not completely known how the curvature changes affect long-term cardiovascular health. Our tortuosity tool allows users to leverage existing aortic models and easily induce different levels of curvature. This supports computational efforts aimed at understanding the effects of tortuosity on hemodynamic quantities, such as wall shear stress \cite{Zhang2021}. For example, previous studies have indicated the potential for tortuosity to negatively affect the performance of stents in thoracic endovascular aortic repair (TEVAR) \cite{Belvroy2019}. Our tool can support researchers in modeling tortuous aortas \textit{in silico} to develop improved stent designs that may ultimately improve the outcomes of TEVAR. To this end, we demonstrate the capabilities of our tool by inducing varying degrees of curvature in a healthy patient-specific aorta. Examples of tortuous shapes created, as inspired by the aortic examples in \cite{Zhang2021}, are shown in Fig.~\ref{fig_tortuosity_examples}. The 3D velocity field and time-averaged wall shear stress (TAWSS) for examples, \textit{t01} and \textit{t04}, are shown in Fig.~\ref{fig_aorta_results}. Higher levels of wall shear stress are seen in the tortuous regions of the descending aorta, compared to the normal case, agreeing with observations reported in literature \cite{Zhang2021}.
    
    \subsection{Creating aneurysms in an aortofemoral model}
    
        Our aneurysm tool allows users to quickly create aneurysms of varying shapes and sizes in existing anatomic models. This is particularly aimed at supporting efforts towards elucidating the hemodynamic variables governing aneurysm rupture, thrombus formation, and vascular growth and remodeling \cite{Rissland2008, Wilson2013b}. To this end, we generate a range of abdominal aortic aneurysms in an aortofemoral anatomy (Fig.~\ref{fig_aneurysm_examples}). We can generate idealized aneurysms, as shown in case \textit{a01}. Through consecutive applications of our tool, we can also create complex aneurysmal shapes, such as the fusiform aneurysm in \textit{a03}. Furthermore, we can apply our aneurysm tool in combination with the tortuosity tool to create lopsided aneurysms, such as example \textit{a04}. This aneurysmal shape was generated by first applying the tortuosity tool to induce slight curvature in the descending aorta and then applying the aneurysm tool on the curved vessel to create a skewed aneurysm. Case \textit{a04} is motivated by the asymmetric abdominal aortic aneurysms found in patients with leg amputations \cite{vollmar1989}. The 3D streamlines and TAWSS, for these two examples, compared against the original healthy model, are visualized in Fig.~\ref{fig_aortofemoral_results}. We observe increased flow recirculation and reduced TAWSS, features typically characteristic of aneurysms.
    
    \subsection{Generating stenoses in coronary arteries}
    
\begin{figure}
        \centering
        \includegraphics[width = 3.25in, height = \textheight, keepaspectratio]{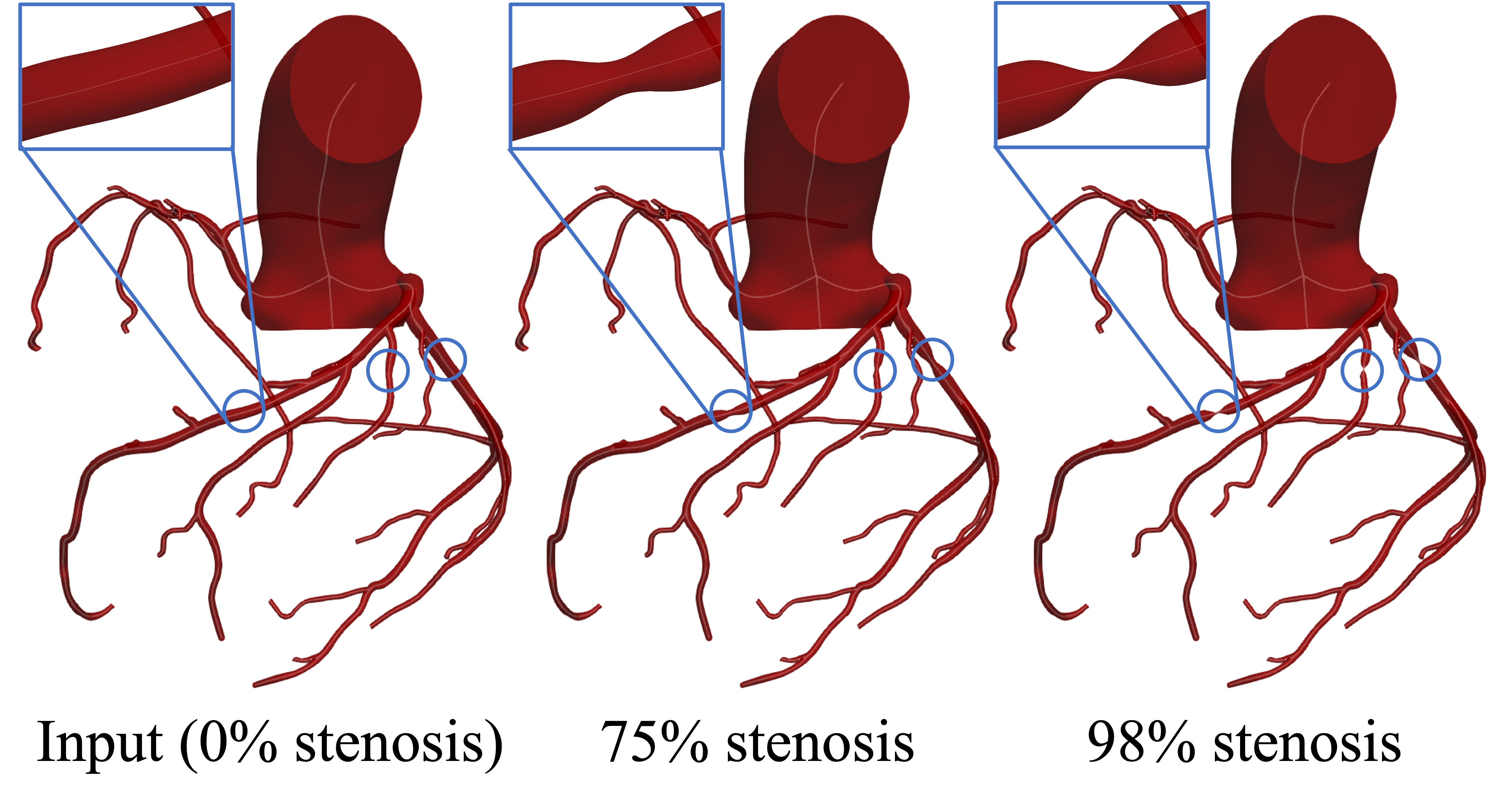}
        \caption{We use our stenosis tool to generate stenoses of varying severities, as defined by a reduction in the cross-sectional area of the lumen, in a patient-specific coronary model. The stenoses are shown in the blue borders. \label{fig_stenosis_examples}}
    \end{figure}

\begin{figure*}
        \centering
        \includegraphics[width = \textwidth, height = \textheight, keepaspectratio]{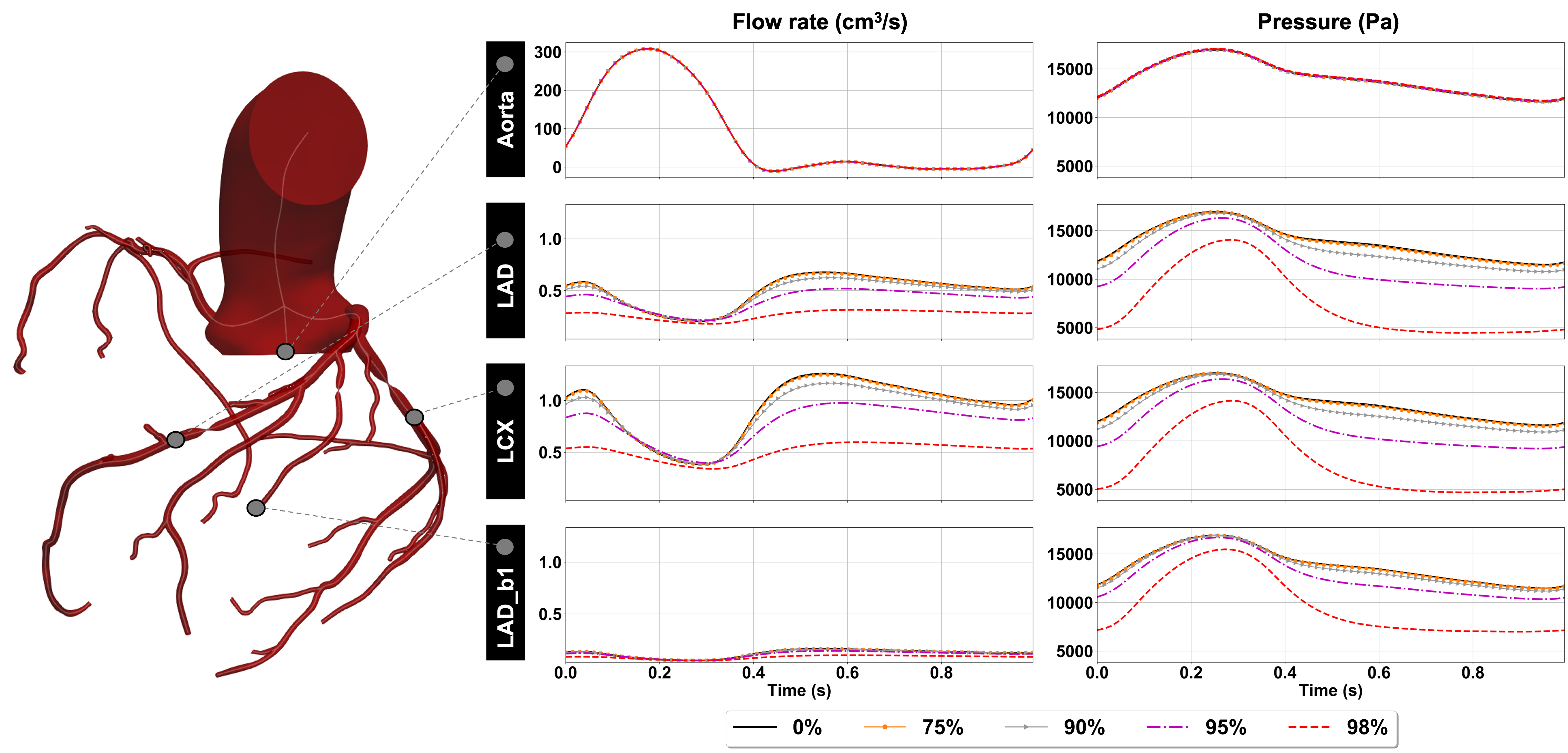}
        \caption{We simulate the stenosed coronary anatomies with 0D models to quickly estimate the impact of the shape changes on the flow rates and pressures. Increasing the constriction in the left anterior descending artery (LAD), one of its daughter branches (LAD\_b1), and the left circumflex artery (LCX) reduces the downstream coronary blood flow and pressure. The systemic circulation, however, is largely unaffected by presence of the stenoses. \label{fig_coronary_results}}
    \end{figure*}

        We apply our stenosis tool to automatically constrict coronary arteries in a patient-specific model and create stenoses of varying levels of severity. We show examples of the generated stenoses in Fig.~\ref{fig_stenosis_examples}. To quickly estimate the hemodynamic impacts resulting from the introduction of the stenoses, we simulate the flows and pressures using 0D models with an expansion pressure loss coefficient in the stenotic regions and coronary boundary conditions \cite{Pfaller2022, kim2010b}. The flow rates and pressures downstream of the stenoses in the coronary arteries, as well as at the inlet of the aorta, are plotted in Fig.~\ref{fig_coronary_results}. The 0D model captures the expected physiologic behavior of coronary hemodynamics, in which pressure and flow waveforms are out-of-phase. As expected, increasing the severity of the stenoses leads to reduced downstream blood flow in the coronary arteries distal to the stenosis.

\section{Discussion}
    In this work, we introduced svMorph, a general-utility framework for interactive morphing of vascular models. Our shape-editing tools enable users to produce a range of altered anatomies relevant to disease states and treatment planning applications. We demonstrated three capabilities of svMorph: tortuosity inducement, aneurysm generation, and stenosis creation. These tools, which can be applied individually or in combination, enable researchers to leverage existing computational models of patient-specific vascular anatomies and quickly deform them for subsequent analysis. This circumvents the need for tedious pathline and segmentation modification processes. Our geometry-manipulation tools also operate directly on the input centerlines and surface mesh, as opposed to the morphMan tool, reducing the need for auxiliary, potentially memory-intensive data structures \cite{Kjeldsberg2019}. Furthermore, all interactions performed in svMorph use just a computer keyboard and a mouse, commonplace equipment regularly available to most vascular simulation practitioners and clinicians, rather than specialized hardware, such as that used in SURGEM \cite{Pekkan2008, Luffel2016}. Unlike other frameworks, svMorph is also integrated with a reduced-order 0D solver, enabling users to quickly estimate the impact of geometric changes on hemodynamics.
    
    We plan to migrate svMorph into the open-source SimVascular software package to make our tools available in the future. However, we note that our morphing tools are not without limitations. One limitation is our tortuosity tool simulates only elastic-like deformation. The OP algorithm exploited is not fundamentally rooted in continuum mechanics. Our tortuosity tool is thus unable to capture Poisson effects, where elongation along the centerline axis would yield contraction in the transverse directions. However, this is an expected result of our tool, as we did not require the deformations to be bound by mechanics. Furthermore, as the primary intended application of this tool is to support model sculpting activities and to reduce the need for manual segmentation-editing, we find the elastic-like deformations of the tortuosity tool to be adequate. In future work, however, we will investigate the use of deformers derived from mechanics-based principles to enable realistic surgical planning operations \textit{in silico}. True elasticity-based deformers, such as the regularized Kelvinlet, are of particular interest \cite{DeGoes2017}.

    Our aneurysm tool also expands the vessel lumen without accounting for potential narrowing from intraluminal thrombus formation. As such, this tool is limited primarily to modeling studies where thrombus is not present. Future work could entail improving the aneurysm tool to account for thrombus-induced lumen narrowing to enable more accurate CFD studies. However, this would require a detailed characterization of thrombotic regions from medical imaging data.
    
    Furthermore, we represent the shape of stenoses in this work using analytical functions, namely, a Gaussian profile. We find this assumption to be suitable for our intended sculpting applications. However, the shape of the stenosis has been shown to play an important role in hemodynamics \cite{Haldar1985, Freidoonimehr2021}. Future studies should comprehensively characterize the geometric features of stenosis shapes from in vivo medical imaging studies. Such work is needed to develop morphing tools that can properly model realistic stenotic geometries.
    
    Additionally, svMorph uses a centerline-based methodology to edit a patient-specific surface mesh. This approach is useful in applications where global shape edits are required. However, our tools are unable to apply local shape edits. To enable more complex surgical morphing operations, we plan to develop methods to perform localized deformations on the surface mesh. We will also include other common vascular shape-editing operations, such as graft placement and stenting.
    
    We also note that our morphing methods are mesh-dependent. The smoothness of the morphed models depends on the mesh resolution of the input mesh. If the input mesh density is coarse, for example, then the morphed geometry may exhibit distortion effects when a severe constriction is generated. This relationship between the input and output mesh qualities is a reasonable limitation, however, as the smoothness of the input model is itself dependent on the mesh density. Any distortion caused by a coarse mesh in the morphing process can be resolved by applying SimVascular's native re-meshing tools. One avenue for future improvement is mesh-independent morphing via implicit surfaces \cite{CaniGascuel1997}. However, this approach require a complex conversion between the input mesh and an analytical function describing the model surface \cite{Botsch2010}. Defining the implicit surface for an arbitrarily shaped vascular network is also non-trivial.
    
    Finally, svMorph employs 0D models with stenosis loss coefficients to rapidly predict the hemodynamic changes induced by radial contractions in blood vessels \cite{Pfaller2022}. However, our 0D models do not include curvature loss coefficients and thus cannot accurately estimate the hemodynamics in such features. Similarly, current 0D models cannot completely capture the geometric effects of expansions and asymmetry present in fusiform and asymmetric aneurysms, thereby preventing 0D models from simulating the effects of common aneurysmal hemodynamic phenomenon, such as recirculation. Future work could expand 0D modeling methodologies to capture these geometric effects. With regards to curvature loss, for example, different loss functions, such as those outlined in the work by Jadhav et al. \cite{Jadhav2017}, can be explored. Improving the accuracy of 0D models would increase the utility of svMorph for virtual surgery planning applications. It would enable clinicians to interactively perform vascular surgical operations and immediately gauge the effects of the geometric changes in a risk-free environment. In realistic surgical planning scenarios, such a tool could potentially lead to improved post-operative outcomes.

\begin{acknowledgment}
    We thank Ingrid Lan, Aekaansh Verma, Zachary Sexton, Suhaas Anbazhakan, and Erica Schwarz for software support and manuscript reviews. The Stanford Research Computing Center provided the computing resources used for our 3D simulations. This work is supported by NIH grants R01EB029362 and R01LM013120.
\end{acknowledgment}

\bibliographystyle{asmems4}

\bibliography{references}

\end{document}